\documentclass[letterpaper, 10 pt, conference]{ieeeconf}  % Comment this line out if you need a4paper

\IEEEoverridecommandlockouts                              % This command is only needed if 
\usepackage{amsmath} % assumes amsmath package installed
\usepackage{amssymb}  % assumes amsmath package installed
\usepackage{fancyvrb, listings, color, float, graphicx, subfigure}
\usepackage{bm}
\usepackage{xcolor}
\usepackage{algorithm}
\usepackage{algpseudocode}
\usepackage{siunitx, booktabs, etoolbox}
\usepackage{graphicx}
\usepackage{wrapfig}
\usepackage{url}

\definecolor{darkgreen}{rgb}{0, 0.5, 0}
\definecolor{myorange}{rgb}{1.0, 0.4, 0}

\makeatletter
\newcommand{\pushright}[1]{\ifmeasuring@#1\else\omit\hfill$\displaystyle#1$\fi\ignorespaces}
\newcommand{\pushleft}[1]{\ifmeasuring@#1\else\omit$\displaystyle#1$\hfill\fi\ignorespaces}
\makeatother

\DeclareMathOperator*{\minimize}{minimize}

\DeclareMathOperator*{\st}{subject\;to}
\def\begm{\ensuremath\begin{bmatrix}}
\def\endm{\ensuremath\end{bmatrix}}

\newtheorem{theorem}{\noindent \normalfont \textbf{Theorem}}

\newcommand{\E}{\mathbb{E}}

\newcommand{\xj}{x^{(j)}} % state at time j
\newcommand{\xjbar}{\bar{x}^{(j)}} % linearization point
\newcommand{\dxj}{\Delta x^{(j)}}
\newcommand{\dxjp}{\Delta x^{(j+1)}}
\newcommand{\xjp}{x^{(j+1)}} % state at time j+1
\newcommand{\xtraj}{\bm{x}}
\newcommand{\dxtraj}{\Delta\bm{x}}
\newcommand{\statespace}{\mathbb{R}^{s}} % state space
\newcommand{\xdist}{\mathcal{X}_0} % distribution over initial state

\newcommand{\Xj}{\mathbb{X}^{(j)}} % state constraint

\newcommand{\uj}{u^{(j)}} % action at time j
\newcommand{\ujbar}{\bar{u}^{(j)}} % linearization point
\newcommand{\duj}{\Delta u^{(j)}}
\newcommand{\utraj}{\bm{u}}
\newcommand{\dutraj}{\Delta\bm{u}}

\newcommand{\oj}{y^{(j)}} % state at time j
\newcommand{\obsvspace}{\mathbb{R}^{o}} % state space

\newcommand{\wj}{w^{(j)}} % state at time j
\newcommand{\vj}{v^{(j)}} % state at time j

\newcommand{\actionspace}{\mathbb{R}^{a}} % action space

\newcommand{\Uj}{\mathbb{U}^{(j)}} % action constraint

\newcommand{\fj}{f} % dynamics at time j
\newcommand{\gj}{g} % dynamics at time j

\newcommand{\costj}{c^{(j)}} % cost at time j

\newcommand{\consensusHorz}{N_\text{c}}

\title{\LARGE \bf
Particle MPC for Uncertain and Learning-Based Control
}

\author{Robert Dyro, James Harrison, Apoorva Sharma, Marco Pavone% <-this % stops a space
\thanks{Authors are with Stanford University, Stanford CA 94305, USA.
        {\tt\small \{rdyro, jharrison, apoorva, pavone\}@stanford.edu}}%
}

\renewcommand{\baselinestretch}{.972}

\begin{document}

\maketitle
\thispagestyle{empty}
\pagestyle{empty}

%%%%%%%%%%%%%%%%%%%%%%%%%%%%%%%%%%%%%%%%%%%%%%%%%%%%%%%%%%%%%%%%%%%%%%%%%%%%%%%%
\begin{abstract}
As robotic systems move from highly structured environments to open worlds, incorporating uncertainty from dynamics learning or state estimation into the control pipeline is essential for robust performance. 
In this paper we present a nonlinear particle model predictive control (PMPC) approach to control under uncertainty, which directly incorporates any particle-based uncertainty representation, such as those common in robotics. 
Our approach builds on scenario methods for MPC, but in contrast to existing approaches, which either constrain all or only the first timestep to share actions across scenarios, we investigate the impact of a \textit{partial consensus horizon}.
Implementing this optimization for nonlinear dynamics by leveraging sequential convex optimization, our approach yields an efficient framework that can be tuned to the particular information gain dynamics of a system to mitigate both over-conservatism and over-optimism.
We investigate our approach for two robotic systems across three problem settings: time-varying, partially observed dynamics; sensing uncertainty; and model-based reinforcement learning, and show that our approach improves performance over baselines in all settings. 

\end{abstract}

%%%%%%%%%%%%%%%%%%%%%%%%%%%%%%%%%%%%%%%%%%%%%%%%%%%%%%%%%%%%%%%%%%%%%%%%%%%%%%%%
%%%%%%%%%%%%%%%%%%%%%%%%%%%%%%%%%%%%%%%%%%%%%%%%%%%%%%%%%%%%%%%%%%%%%%%%%%%%%%%%
\section{Introduction}

As autonomous decision-making agents such as robots move from narrowly tailored environments and begin operating in unstructured, uncertain worlds, incorporation of uncertainty into the decision-making pipeline is essential to intelligently trade off risk and reward. Uncertainty is ubiquitous in robotic systems; from state estimation, to fault detection, to dynamics learning, approximate Bayesian estimation and filtering methods hold central roles throughout the autonomy stack. Moreover, as robots regularly enter novel environments, interact with humans, or perform novel tasks, consideration of uncertainty  is becoming increasingly consequential.

% Incorporating uncertainty into the decision-making pipeline is necessary to ensure systems are accurately considering the full distribution of possible outcomes when choosing actions. Despite the benefits of this probabilistic reasoning, \textit{certainty equivalent} methods---in which a point estimate is used in downstream control---still dominate. Outside of these methods, \textit{robust} methods, which consider adversarial disturbances at each time step, are common. However, these methods are typically over-conservative due to their worst-case, set theoretic approach. In this work, we investigate particle-based model predictive control that enables tunable conservatism and pairs naturally with Bayesian models common in state estimation and dynamics learning such as sequential (particle filter) Monte Carlo and ensemble neural networks. Our approach is based on a discrete particle representation of distributions and is thus applicable to arbitrary representations of uncertainty. 

% Current approach to uncertainty in robotics
Modern robotic systems already quantify uncertainty in the state-estimation or dynamics learning process, through e.g. particle filtering or ensemble learning. Nevertheless, this uncertainty is often ignored in action selection: \textit{certainty equivalent} control methods -- which ignore uncertainty when choosing actions -- are commonplace due to their ease of implementation, computational efficiency, and practical performance. 
% Existing approach to uncertainty in control
Within the controls community, there exist several techniques for factoring in state and model uncertainty into an optimal control framework, yet these methods are often too conservative (failing to account for future information gain and/or assuming adversarial disturbances) \cite{blackmore_probabilistic_2010} or too computationally intensive, limiting their applicability to systems with slower dynamics, such as process control \cite{lucia2013multi}.

% Where we stand
In this work, we build on uncertainty-aware optimal control approaches developed in the controls community, specifically scenario-based model predictive control (MPC), which optimizes a sequence of actions for sampled possible realizations of the uncertainty. 
Notably, in contrast to existing work which either constrains actions to be shared across scenarios for either the \textit{full} planning horizon (leading to over-conservatism), or only the first time step (leading to over-optimism), we employ a \textit{tunable consensus horizon} allowing a practitioner to tailor the controller to the information gain dynamics of their application. Our approach operates directly on the particle-based uncertainty representations commonplace in robotic systems, and, leveraging sequential convex programming, handles nonlinear dynamics while remaining computationally efficient. 

\subsection{Related Work}
The problem of optimal control with uncertainty is the focus of the fields of stochastic and robust optimal control, each of which have long histories. A common approach is the class of robust methods---such as methods from classical robust control---which factor in uncertainty via assuming worst-case disturbances \cite{kothare_robust_1996, khargonekar_hsub_1988}.
However, if certain disturbances are less likely than others, this worst-case treatment of uncertainty can lead to overly conservative behavior, limiting practical use in robotic systems. 

To handle probabilistic disturbances, work over the last several decades has focused on distributional approximations of uncertainties, typically approximating the disturbance distribution as a Gaussian. While these approaches, such as those of \cite{cinquemani_convexity_2011, deisenroth2011pilco, deisenroth2012learning}, yield tractable solutions, the Gaussian approximation can be poor for the arbitrary, potentially multi-modal uncertainties that are commonplace in robotics, e.g. a chance of actuator failure. 
Moreover, works which model the uncertainty as additive i.i.d. noise, such as \cite{kantas_sequential_2009, deisenroth2011pilco}, ignore the temporal correlation of uncertainty. This is particularly problematic for dynamics uncertainty: if an unknown dynamics parameter is non-time-varying, the impact of this parameter will be different (and typically result in a wider distribution of outcomes) than if the same parameter is re-sampled at every time step \cite{mchutchon2015nonlinear}. However, consideration of temporal correlation in distributional approximations typically leads to computational difficulties. 

Difficulties in handling temporal correlation of disturbances has motivated sampling-based (or similarly, particle methods or scenario methods) approaches to handling uncertainty. Several works have considered the case of linear systems, which yields a large convex optimization problem composed of all the scenarios \cite{blackmore_probabilistic_2010, calafiore_robust_2012}. In the nonlinear setting, iterative importance sampling control schemes such as that of \cite{williams2017information} have seen application in settings with model uncertainty \cite{chua_deep_2018, abraham2020model, arruda2017uncertainty}. These methods do not rely on the gradient of the cost function (or dynamics) for optimization, and are strictly sampling-based. 
In contrast to these works, our framework is capable of handling arbitrary nonlinear systems, and uses available gradients to improve performance, as we demonstrate in our experiments.  

Another family of approaches is multi-stage MPC, which consider a branching tree structure (corresponding to sampled scenarios or disturbances). Rather than share actions across all branches, as in robust control methods, multi-stage MPC imposes causality constraints by sharing actions only at each branch point, thus preventing the controller from anticipating a disturbance before it occurs \cite{lucia2013multi}. The tree structure induced by these approaches has lead multi-stage methods to consider both a ``robust'' phase, in which branching occurs, and a tail phase in the control optimization in which branching no longer occurs and thus causality constraints are not imposed. While this approach is more reflective of the true information pattern available to the controller during online operation, it is either expensive due to an explosion in the number of sampled scenarios, or under-conservative \cite{lucia2014handling}. Indeed, the computational complexity of these multi-stage methods has so far restricted their application to problems with slow dynamics, such as process control \cite{lucia2014handling, marti2015improving}. Moreover, in the partially observed setting, reflecting the true information pattern requires either (differentiably) simulating the state estimator forward in time \cite{thangavel2018dual, arcari2019approximate}, or may result in substantial under-conservatism due to assumed perfect state information after a small number of steps. These limitations stand in stark contrast to methods such as \cite{williams2017information} that have seen widespread use in learning-based control due to their simplicity and broad applicability. In contrast to the literature on multi-stage MPC, we enforce a constraint requiring matching actions for an extended horizon while limiting the branching of the scenario tree, resulting in a simple, tunable, and effective approach.

\subsection{Contributions}
In this work, we propose a particle model predictive control (PMPC) formulation that leverages sequential convex programming (SCP) and a tunable consensus horizon to yield an efficient nonlinear control framework capable of handling arbitrary uncertainties. 
This particle MPC approach, in contrast to multi-stage or scenario MPC methods, minimizes branching and focuses on principled simplifications to make the approach computationally efficient, tunably conservative, and simple to implement. 
We argue that this formulation, based on Monte Carlo distributional approximations as opposed to methods based on set theoretic uncertainty or approximation with analytically tractable distributions, is a highly flexible and effective approach to control under uncertainty. 
We investigate the role of the consensus horizon both theoretically---showing that its optimal choice depends on the information gain rate in the system---as well as experimentally.

We investigate three problem settings: learning-based control (or equivalently, model-based reinforcement learning) in which an agents learns a model of the system dynamics with no prior information; time-varying parametric uncertainty such as faults or wind gusts; and sensing uncertainties such as those arising from particle filter state estimation. All three problems are investigated on two different systems---a free-flying space robot system and a planar quadrotor---and 
we find significant performance gains from our approach relative to the standard certainty equivalent and uncertainty-aware sampling-based control schemes.

%%%%%%%%%%%%%%%%%%%%%%%%%%%%%%%%%%%%%%%%%%%%%%%%%%%%%%%%%%%%%%%%%%%%%%%%%%%%%%%%
\section{Problem Formulation}

Our goal in this work is to control partially observed dynamical systems. This general formulation, a Partially Observable Markov Decision Process (POMDP), captures several sources of uncertainty in robotics. We denote the system state at time step $j$ as $\xj \in \statespace$, the action taken from this state $\uj \in \actionspace$, and the observation as $\oj \in \obsvspace$. 
% We will assume throughout this work that only the observations are observed. 
The system has nonlinear, discrete time dynamics
\begin{align*}
    \xjp = \fj(\xj, \uj, \wj), \quad\quad
    \oj = \gj(\xj, \vj)
\end{align*}
where $\fj(\cdot)$ and $\gj(\cdot)$ are the state dynamics and observation function respectively, and $\wj, \vj$ are stochastic disturbances.

As solving POMDPs directly is intractable \cite{wittenmark1995adaptive}, in robotic systems this problem is typically decomposed into \textit{state estimation}, which uses observations to estimate the current state $\xj$, and \textit{control}, which aims to select the action $\uj$ given this estimate. 
The state can not be known exactly, robotic systems often turn to Bayesian filtering to maintain an update a \textit{belief distribution} $\xdist$ over possible states. Note that this formulation can also characterize uncertain dynamics, where the unknown parameters of the dynamics can be considered as unobserved state elements which remain constant over time. 

Our focus is control: choosing an optimal action to take given the current belief over state. 
Given information about the control task in the form of a stage-wise cost function
$\costj(\xj,\uj)$ and state and action constraints $\xj \in \Xj$, $\uj \in
\Uj$, this gives rise to the finite horizon optimal control problem
\begin{equation}
\label{eqn:fhoptcontrol}
\begin{split}
\minimize_{\utraj}~~& \sum_{j=0}^N \E_{\xj}\left[\costj(\xj, \uj) \right]\\
\st~~& \xjp = \fj(\xj,\uj,\wj) \\
 & \xj \in \Xj \,\, \forall j\\
 & \uj \in \Uj \,\, \forall j\\
 & \wj \sim \mathcal{W}^{(j)} \,\, \forall j\\
 & x^{(0)} \sim \xdist
\end{split}
\end{equation}

We consider the cases where $\xdist$ is a discrete distribution over possible values of $x^{(0)}$, as is the case with a particle filtering approach to state estimation, or an ensemble or sampling based approach to quantifying uncertainty in learned dynamics models. Note that if the true system is not in the support of this discrete distribution, then guaranteeing (even probabilisitically) that the true system will also satisfy state constraints is an open research question which is outside the scope of this paper.

%%%%%%%%%%%%%%%%%%%%%%%%%%%%%%%%%%%%%%%%%%%%%%%%%%%%%%%%%%%%%%%%%%%%%%%%%%%%%%%%
\section{Particle Model Predictive Control}

We propose to solve the optimal control problem \eqref{eqn:fhoptcontrol} using a receding horizon approach; at each time step, we optimize a sequence of $N$ control actions into the future. By replanning at every step, we are able to account for the evolution of the state estimate, which changes as new observations are made. 
Receding horizon control is a well established technique in applications of MPC and we make the approach explicit in Algorithm \ref{alg:rh_control}.
% We propose to approach our problem using the typical receding horizon approach with replanning at every step. 

The fundamental challenges in solving \eqref{eqn:fhoptcontrol} comes from (i) the presence of uncertainty and (ii) the nonlinear dynamics. In this section, we detail our algorithmic approach, first focusing on how we handle uncertainty, and then discussing how we deal with the nonlinear dynamics. 

\subsection{Handling Uncertainty through Particle Model Predictive Control}

In designing a control strategy for this problem, we need an approach which can factor in the temporally correlated effects of state uncertainty. At the same time we need an approach which remains computationally efficient, so that we can replan online to handle deviations due to the inevitable plan mismatch with reality and factor in online updates in state belief updates coming from online estimation. In order to do so, we propose to operate directly on a particle-based representation of state uncertainty.

Formally, our particle model predictive control (PMPC) approach has the structure of the finite horizon optimal control problem \eqref{eqn:fhoptcontrol}. We approximate the uncertainty in the initial state and dynamics by a collection of particles. Each particle represents the evolution of the system under one realization of the initial state. That is, for particle $i$, the state evolves according to $\xjp_i = \fj(\xj_i, \uj_i, \wj_i)$, where $x^{(0)}_i \sim \xdist$.
By propagating each particle separately in time, we account for the time correlated effects of dynamics and state uncertainty in contrast to approaches which handle uncertainty through uncorrelated disturbances at each time step. Importantly, for each particle we get a \textit{separate trajectory} with potentially different states and control actions. 

\begin{algorithm}[t]
\caption{Receding horizon PMPC with learning}
\label{alg:rh_control}
\begin{algorithmic}[1]
\Require{State belief $\mathcal{X}_0$}
\State Sample $\{x^{(0)}_i\}_{i=1}^M, \{w_i^{(j)}\}_{i=1}^M \sim (\mathcal{X}_0, \mathcal{W}_j) $  particles and disturbances from current belief (state, unknown disturbances)
\State Compute the PMPC plan $\{\xtraj_i, \utraj_i \}$ on sampled particles. 
\State Execute first action of plan $u^{(j=0)}$.
\State Observe transition $y \gets g\big(f(x = x^{(0)}, u^{(j=0)}, w\big)$
\State $\mathcal{X}^{(0)} \gets $ Updated belief after observing $(y, u^{(j=0)})$
\end{algorithmic}
% \vspace*{0.1cm}
\end{algorithm}

For applications in real systems, we employ the well known strategy of receding horizon control with replanning at every step. The overview of our approach is contained in Algorithm \ref{alg:rh_control}. We approximate the expectation in \eqref{eqn:fhoptcontrol} via a (potentially weighted) sum of costs across these particles. 

We enforce constraints uniformly over all sampled particles. This is only guaranteed to ensure the true system satisfies constraints if the set of particles includes the true system state and noise realization; if this is not the case then guaranteeing constraint satisfaction (even at some probability) is in general a hard set-theoretic problem in stochastic control, especially for arbitrary uncertainty distributions. In our experiments, we found that with a reasonable number of particles, our approach was sufficient, but developing practical guarantees on probabilistic constraint satisfaction remains an important direction of research.
In our implementation, we handle state constraints with exact penalty functions \cite{kerrigan_soft_2000}, which ensures constraint satisfaction when the problem is feasible, and minimizes violation if the problem becomes infeasible.

Even as we simulate \textit{several} possible future trajectories, or scenarios, in this optimization, we must choose a \textit{single} action to execute.\footnote{
Even though some works suggest combining MPC with a feedback policy \cite{ono_closed-loop_2012, ridderhof_nonlinear_uncertainty_2019}, computed on- or offline, we experimentally found that to be computationally challenging in real-time applications and yielding little improvement in receding horizon control. Hence, we do not consider a feedback policy in our MPC problem.} This introduces a key question of how to enforce \textit{consensus} among the different scenarios.
Existing approaches have focused either on \textit{full consensus}, where we ensure actions are shared across particles over all time steps ($\uj_i = \uj, \forall i=1,\dots,M,~~j=0,\dots,N$), or \textit{one-step consensus} where only the first action (the one that will be executed) is shared across scenarios ($u^{(0)}_i = u^{(0), \forall i=1,\dots,M}$). Full consensus ensures that actions perform well---open-loop---across all sampled scenarios. However, as robotic systems employ filtering techniques which reduce uncertainty at each timestep, this procedure can be over-conservative, and impractical for robotics applications. One-step consensus, on the other hand, allows choosing actions tailored to each scenario after the first time-step, and thus may be over-optimistic, and thus lead to suboptimal performance on the true uncertain system.
A key contribution of this work is to consider interpolating between these two extremes, which each have their advantages and drawbacks, through a \textit{tunable consensus horizon}. Specifically, we define the consensus horizon $\textcolor{red}{\consensusHorz}$, as the number of timesteps for which actions must be shared across particles ($u^{(j)}_i = u^{(j)}_k ~~~ \forall j \in [0..\textcolor{red}{\consensusHorz}] ~~ i, k \in [1..M]$). We discuss the implications of this hyperparameter further in Section \ref{sec:consensusHorz}.

\subsection{Combining Particle Representations with Sequential Convex Programming}

The PMPC approach yields a large non-convex optimization problem. In order to approximately solve this problem efficiently, we employ sequential convex programming (SCP). SCP applies efficient, optimized solvers for convex optimization problems sequentially to locally optimize a nonlinear optimization problem. At each step of SCP, we first form a convex approximation of the non-convex problem around a reference setting of the optimization variables, and then solve this convex problem to get a new reference point---while attempting to stay close to the previous reference point. SCP can be shown to converge to a local optimum under some mild assumptions, for further discussion we refer the reader to \cite{dinh_local_scp_2010}. 

In the case of PMPC, the optimization variables are the state and action
trajectories of each particle $\{ \xtraj_i, \utraj_i \}_{i=1}^M$ where $\xtraj_i = (x^{(0)}_i, \dots, x^{(N+1)}_i)$ and $\utraj_i = (u^{(0)}_i, \dots, u^{(N)}_i)$. To convexify the problem at a particular setting of these optimization variables, we replace the dynamics with a linear approximation and the cost with a linear approximation of its non-convex part, yielding the convex optimization problem:
\begin{equation}
\label{eqn:convexified_pmpc}
\begin{aligned}
\min_{\{ \dxtraj_i, \dutraj_i \}_{i=1}^M} & \frac{1}{M} \sum_{i=1}^M
\sum_{j=0}^N  \bigg( c_\text{cvx}(\xj_i + \dxj_i, \uj_i + \duj_i) \\
& \pushright{+ \nabla_x \costj_\text{ncvx} \dxj_i + \nabla_u \costj_\text{ncvx} \duj_i} \\
& \pushright{+ \rho_x\| \dxj_i \|_2^2 + \rho_u\|\duj_i\|_2^2 \bigg)} \\
\mathrm{s.t.}~~& \dxjp_i = \nabla_x \fj_{i} \dxj_i + \nabla_u \fj_{i} \duj_i \quad % \\ &\quad\quad 
\\
  & \pushright{\forall j \in [0..N]~~ i \in [1..M]} \\
 & \xjbar_i + \dxj_i \in \Xj \quad % \\ &\quad\quad 
  \forall j \in [0..N]~~ i \in [1..M]\\
 & \ujbar_i + \duj_i \in \Uj \quad% \\ &\quad\quad 
 \forall j \in [0..N]~~i \in [1..M]\\
 &
 \underbrace{\duj_i = \duj_k \quad \forall j \in [0..\textcolor{red}{\consensusHorz}]~~ i, k \in [1..M]}_{\text{partial consensus}}
\end{aligned}
\end{equation}
where this optimization problem is written in terms of the deviations $\{
\dxtraj_i, \dutraj_i \}_{i=1}^M$ from the previous solution $\{ \bar{\xtraj}_i,
\bar{\utraj}_i \}_{i=1}^M$, and the gradients are evaluated at $\{
\bar{\xtraj}_i, \bar{\utraj}_i \}_{i=1}^M$. 
Note that we add two terms to the cost to minimize deviation from the
linearization point, as the approximation is only good for small deviations.
While there are many strategies for limiting this deviation in each SCP
iteration, we choose the quadratic penalties as they work well in practice and
introduce only two hyperparameters $\rho_x$ and $\rho_u$. We find that the algorithm is generally not sensitive to the particular value of these hyperparamters, as long as they are above a certain threshold: Larger values of $\rho_x, \rho_u$ slow the convergence of the SCP procedure, but too low a value leads to instability of the SCP procedure. 
In practice we assume a multiplicative relation between the two, i.e., $\rho_x = \alpha \rho_u$ and search for a good value $\rho_x$ using a few steps of bisection. Both hyperparameters affect the optimization in the same manner---they penalize the variable deviation from the linearized trajectory---which justifies reducing them to a single hyperparameter. Explicitly recognizing separate penalization of state and control action deviation allows the system designer to account for difference in the order of magnitude of state and control actions, in cases where, for examples, control actions are confined between 0 and 1 and state between -100 and 100. 
SCP requires an initial trajectory guess, and choosing this initialization remains an open research problem \cite{bonalli_gusto_2019}. In our experiments, we find that PMPC is robust to the initial trajectory guess; even simple infeasible guesses---such as static trajectory (repetition of the current state belief) and linear interpolation between initial and goal state---work well.

Algorithm \ref{alg:scp_pmpc} details the procedure for one planning iteration. Each iteration takes as input a set of particles, where each particle $i$ has its own initial state, $x^{(0)}_i$, and dynamics deviation $w^{(j)}_i$. Note that in the case of a non-uniform belief over the particles, the relative weighting of the particles can be incorporated by scaling the particle's cost function by that weighting. SCP alternates between convexification of the dynamics and cost functions, and subsequently solving the convex problem \eqref{eqn:convexified_pmpc} by leveraging an off-the-shelf efficient convex solver. If the algorithm converges, we are guaranteed to have a locally near-optimal solution. We detect convergence by evaluating the norm of the change in the solution trajectory, as at a locally optimal solution, solving \eqref{eqn:convexified_pmpc} would yield a solution with deviations equal to zero. When stopped prior to convergence, the solution is approximate up to the error in the linear approximation of the dynamics and the cost.

\begin{algorithm}[t]
\caption{SCP PMPC}
\label{alg:scp_pmpc}
\begin{algorithmic}[1]

\Require{Initial states $\{ x^{(0)}_i \}_{i=1}^M$, dynamics models $\{ f_i
\}_{i=1}^M$, solution guess $\{ \xtraj_i, \utraj_i \}_{i=1}^M$ }
\Require{Hyperparameters $\rho_x, \rho_u, \consensusHorz$}
\Require{Solution tolerance $\epsilon$}
\Repeat
\State $\{ \bar{f}^{(j)}_i, \nabla_x \fj_i, \nabla_u \fj_i \}_{i=1}^M \gets$ Linearize dynamics around the trajectory guess $\{ \xtraj_i, \utraj_i \}_{i=1}^M$\;
\State Split the cost
into the convex and non-convex parts $\{\costj_{i, \text{cvx}}, \costj_{i, \text{ncvx}} \}_{i=1}^M$\;
\State $\{ \nabla_x \costj_{i, \text{ncvx}},\, \nabla_u \costj_{i, \text{ncvx}} \}_{i=1}^M
\gets$ Linearize non-convex cost around $\{ \xtraj_i, \utraj_i
\}_{i=1}^M$\;
\State $\{ \dxj_i, \duj_i \}_{i=1}^M \gets$ Solution to convex prob. \eqref{eqn:convexified_pmpc}

\State $\{ \xtraj_i, \utraj_i \}_{i=1}^M \gets \{ \xtraj_i + \Delta \xtraj_i, \utraj_i + \Delta \utraj_i \}_{i=1}^M$ %update solution trajectory \;
\Until{$\sum_{i=1}^M \sum_{j=0}^N \| \dxj_i \| + \| \duj_i \| < \epsilon$}
%\KwOut{solution trajectory $\{ \xtraj_i, \utraj_i \}_{i=1}^M$}
\State \Return $\{ \xtraj_i, \utraj_i \}_{i=1}^M$
\end{algorithmic}
\vspace*{0.1cm}
\end{algorithm}

%%%%%%%%%%%%%%%%%%%%%%%%%%%%%%%%%%%%%%%%%%%%%%%%%%%%%%%%%%%%%%%%%%%%%%%%%%%%%%%%
\section{Discussion}

\subsection{Choice of Consensus Horizon}
\label{sec:consensusHorz}
A key hyperparameter in PMPC is the consensus horizon $\consensusHorz$, and the range from \emph{one-step} to \emph{full} consensus has a large impact on the closed loop performance of the receding horizon planning controller. 
One way to interpret the choice of consensus horizon is as an approximation of the information gain dynamics: uncertainty is present for $\consensusHorz$ steps, after which it drops immediately to zero when full information is revealed. By enforcing control consensus for $\consensusHorz$ steps, PMPC is assuming that the uncertainty will remain constant for this time period, and so must choose actions that work for all possibilities. After the consensus horizon, controls can be tailored to each particle, which is only possible when the true system is revealed.
Much of the controls and decision making literature has focused on the two extremes for consensus horizon: 
one-step consensus corresponds to the QMDP approximation of POMDPs \cite{littman1995learning}, and full consensus corresponds to robust control or scenario MPC. 
Instead, PMPC allows choosing any consensus horizon between these two extremes, interpolating between these two approaches.
Full consensus chooses one sequence of actions which satisfy constraints across all possible values of the uncertain state. While this ensures safety, it can lead to \textit{over-conservative} solutions which fail to account for future information gain through the state estimation module of a robotic system.
However, choosing a consensus horizon that is too short can lead to \textit{over-optimism}, and result in actions that lead to constraint violations or large costs, as shown in the following theorem:
\begin{theorem}[Consensus horizon \& replanning]
\label{thm:consensus}
For any integers $N^\star \in [2..\infty)$, $K \in [1..N^\star-1]$, there exists an optimal control problem \eqref{eqn:fhoptcontrol} defined by $\xdist,f,c,\mathbb{X},\mathbb{U},\mathbb{W}$ for which running PMPC with a consensus horizon of $\consensusHorz = K < N^\star$ leads to constraint violation, but a consensus horizon of $\consensusHorz = N^\star$ does not.

\end{theorem}

The proof is available in the appendix, available at \cite{dyro2020particle}.

As this theorem shows, even though only the first action in the receding horizon plan is ever executed on the real system, limiting consensus to only this first action can lead to constraint violation and otherwise unsafe closed-loop behavior. This highlights the utility of choosing an intermediate value for the consensus horizon.

Our approach is similar in spirit to choosing a shorter robust horizon is multi-stage MPC \cite{lucia2013multi}, but there are several key differences. First, multi-stage MPC builds a scenario tree, branching at each timestep up to the robust horizon. Actions are only shared at branch points of the tree. In contrast PMPC creates a scenario tree that only branches at the root, but enforces control consensus over an tunable horizon $\consensusHorz$.
Additionally, multi-stage MPC typically incorporates information gain over the robust horizon, in order to determine how to sample uncertainties and branch the tree. 
While factoring in information gain dynamics into action selection leads to closer-to-optimal performance, it comes at a steep computational expense. In multi-stage MPC, this adds a significant computational burden and the robust horizon is often set to 1 for computational efficiency. Indeed, considering this over the full planning horizon yields the \textit{dual control} problem, which is computationally difficult for general nonlinear systems \cite{wittenmark1995adaptive, feldbaum1960dual}. In contrast, we sacrifice optimality and avoid the computationally expensive process of propagating belief dynamics.
Instead, we adopt the simpler information gain dynamics of $\consensusHorz$-step consensus, enabling us to reap the safety benefits of choosing a consensus horizon greater than 1 without suffering a large computational burden associated with belief propagation.

\subsection{Particle Model Predictive Control vs Certainty Equivalent Model
Predictive Control}
A baseline for control under uncertainty is to assume \textit{certainty equivalence} (CE), i.e. choose a set of actions assuming a nominal value for $x^{(0)}$ is exactly correct. With the exception of systems with linear dynamics and Gaussian state uncertainty \cite{bertsekas1995dynamic}, this technique is sub-optimal, but nevertheless is easy to implement and commonly used in robotics. 
CE control exists as a special case of PMPC where we use only a single particle corresponding to the nominal setting of all uncertain quantities. 
CE control is appealing as a simpler, easier to implement solution, especially for low, unimodal uncertainty for which choosing a nominal value for $x^{(0)}$ it is significantly complicated when dealing with multimodal uncertainties for which the mean may not be an accurate representation. For example, with multimodal sensing uncertainty, planning assuming the robot is at the mean state may lead to actions that are unsuitable for all possible scenarios. 
PMPC allows directly translating arbitrary uncertainty  representations on state, dynamics, and cost function to the controller, sidestepping the often complex decision of reducing an uncertain system to a certain one for CE control.
In our experiments, we demonstrate that PMPC outperforms CE control on common sources of uncertainty, both uni- and multi-modal.

%%%%%%%%%%%%%%%%%%%%%%%%%%%%%%%%%%%%%%%%%%%%%%%%%%%%%%%%%%%%%%%%%%%%%%%%%%%%%%%%

%%%%%%%%%%%%%%%%%%%%%%%%%%%%%%%%%%%%%%%%%%%%%%%%%%%%%%%%%%%%%%%%%%%%%%%%%%%%%%%%

\section{Experiments}
% We have presented SCP PMPC as a general purpose control strategy for the types of uncertainty common in robotics systems. To demonstrate the efficacy of this methods, 

We demonstrate the utility of PMPC by evaluating its performance on two nonlinear systems in three settings each, chosen to highlight distinct sources of uncertainty arising in robotics problems. Specifically, we use (1) a 6-D planar quadrotor (2-D action space), a common system for benchmarking highly dynamic control and reinforcement learning algorithms \cite{ivanovic2019barc, gillula2010design, singh2017robust} and (2) a 6-D planar free-flyer, a free floating spacecraft constrained to a plane with actuation through gas thrusters and a reaction wheel (9-D action space) \cite{lew2020safe}. We use first principles continuous dynamics discretized with explicit Runge-Kutta 4 method.

\begin{figure}
\centering
\subfigure[Free-flyer]{
\includegraphics[width=0.22\textwidth]{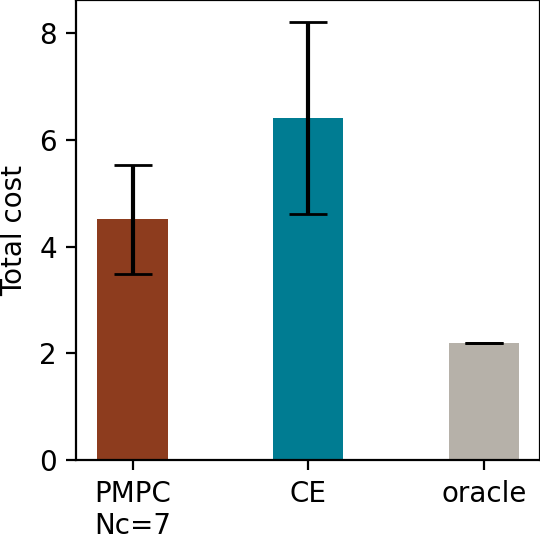}
}
\subfigure[Free-flyer]{
\includegraphics[width=0.22\textwidth]{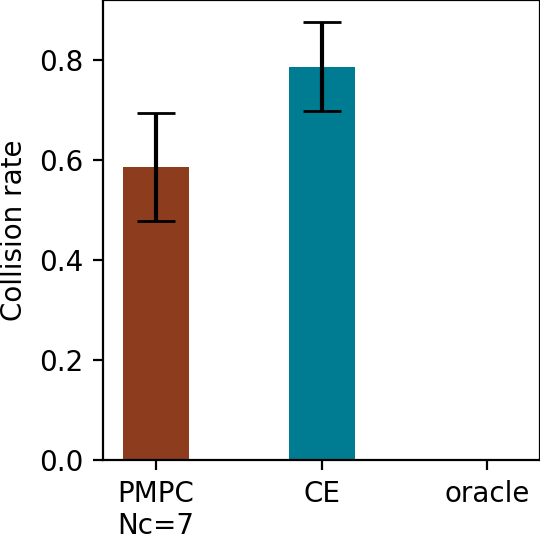}
}
\subfigure[Planar quadrotor]{
\includegraphics[width=0.22\textwidth]{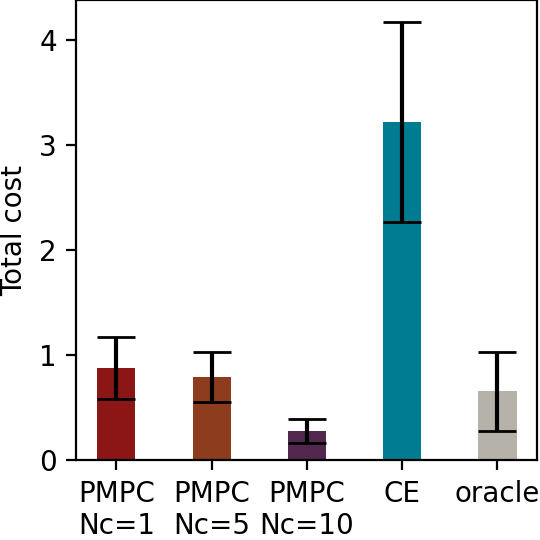}
}
\subfigure[Planar quadrotor]{
\includegraphics[width=0.22\textwidth]{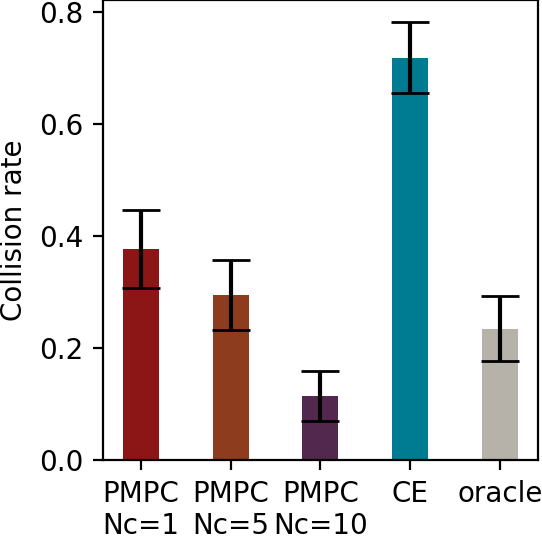}
}
\caption{Total cost and collision rate for the free-flyer and the planar quadrotor under \textbf{model uncertainty}.}
\label{fig:dynamics}
\end{figure}

For each of these systems, we consider three sources of uncertainty that are prevalent throughout robot autonomy: (i) changing system dynamics within an otherwise known dynamics model (e.g. actuator failures or external effects), (ii) state uncertainty, and (iii) epistemic uncertainty within black-box learning-based control (model-based reinforcement learning). 
In each of these settings, we execute a task corresponding to motion to a desired position while avoiding obstacles. 
We encode this task using a cost function which generally takes the form of a quadratic cost on position to encourage movement towards the goal, with exact penalty for collision with obstacles which is zero clipped linear in the amount of violation. Note that the non-convexity of the free space makes this cost function non-convex. We also include experiments investigating the computational complexity of the method for varying numbers of particles, showing the approach is feasible in real time. 

Because of the wide variety of test environments, we exclude details of each environment from the body of the paper. The details for all environments, including details of the design of the particle dynamics in the control scheme, are available in the appendix \cite{dyro2020particle}. This appendix also includes additional experimental results. Throughout all figures in this section, we show 95\% confidence intervals. 

We solve the SCP problem using the open-source QP solver OSQP \cite{stellato_osqp_2020}. Except for dynamic uncertainty Free-flyer with $M=16$, we choose $M=10$ particles.

\subsection{Dynamics Uncertainty}
Dynamics uncertainty is commonplace in robotics. For one, robotic systems may evolve stochastically over time, e.g. having a nonzero chance of actuators failing. They can also be affected by similarly evolving external factors, e.g. wind disturbances. We use the two systems to investigate the performance of PMPC in these characteristic settings of dynamics uncertainty. 

For the free-flyer problem, we consider a scenario in which the robot aims to regulate to a position near a wall, mimicking a docking scenario. For this system, each thruster has a random failure probability at each time step. The current failure status (a discrete random variable) is inferred online by a recursive Bayes filter. We compare to a certainty equivalent (CE) formulation that acts with respect to the maximum a posteriori (MAP) failure state estimate. 

For the planar quadrotor, we assume a wind disturbance modeled by an Ornstein-Uhlenbeck process in both spatial dimensions. This process is, roughly, a Brownian motion process with a term that pulls the process toward 0 and thus captures the stochastic but temporally correlated nature of wind gusts. In this problem, the quadrotor aims to navigate a narrow passage while avoiding collision. The CE baseline assumes the current wind remains constant. 

Results for both systems can be seen in Figure \ref{fig:dynamics}. In both experiments, we also compare to an oracle baseline which has perfect knowledge of future dynamics changes. For the free-flyer system, we plot only $\consensusHorz=7$ as results were consistent across varying consensus horizons. For the planar quadrotor, we visualize performance for $\consensusHorz = 1, 5, 10$. In our experiments, all the PMPC approaches outperformed the CE approach, and avoided collisions substantially more frequently. Interestingly, PMPC with $\consensusHorz=10$ actually outperformed the oracle model. While the oracle model has perfect knowledge of the time evolution of parameters governing stochastic disturbances \emph{within the finite planning horizon}, it fails to account for sample values beyond it. Thus, the conservatism added by the consensus horizon improves performance due to additional robustness to these stochastic disturbances.

\begin{figure}
\centering
\subfigure[Free-flyer]{
\includegraphics[width=0.22\textwidth]{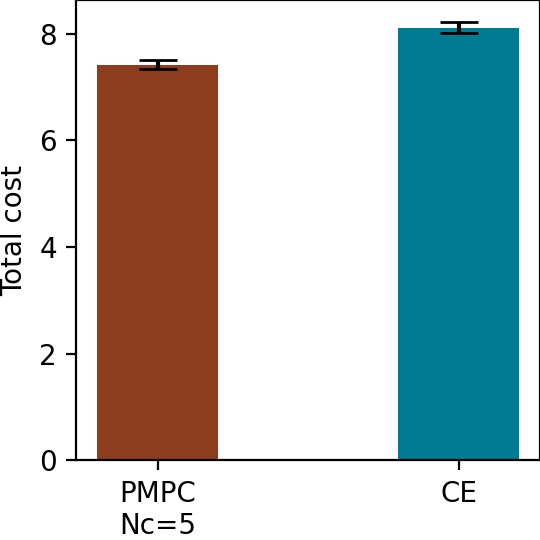}
}
\subfigure[Free-flyer]{
\includegraphics[width=0.22\textwidth]{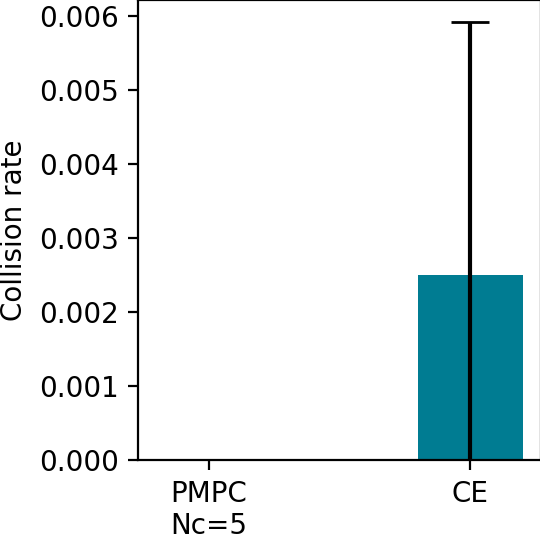}
}
\subfigure[Planar quadrotor]{
\includegraphics[width=0.22\textwidth]{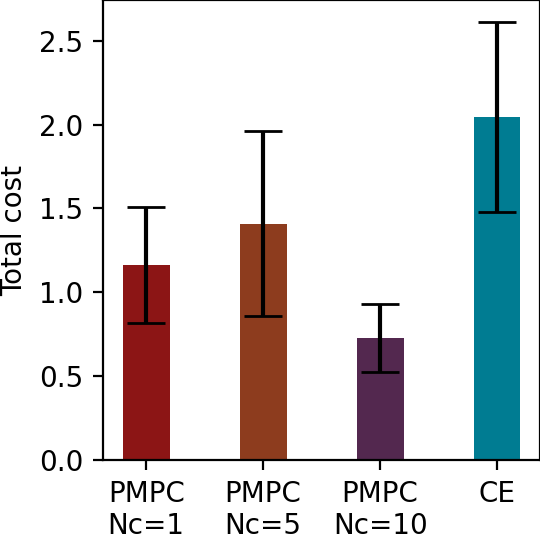}
}
\subfigure[Planar quadrotor]{
\includegraphics[width=0.22\textwidth]{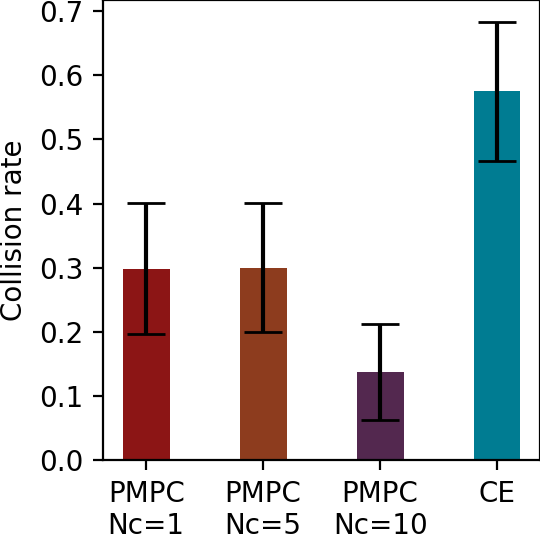}
}
\caption{Total cost and collision rate for the free-flyer and the planar quadrotor under \textbf{state uncertainty}.}
\label{fig:sensing}
\end{figure}

\subsection{State Uncertainty}

A common source of uncertainty in robotics comes from partial observability of the system's state. Typically, robotic systems employ online filtering to estimate their current state, for example, by using a particle filter \cite{thrun2006probabilistic}. We test how well SCP PMPC can address this form of uncertainty on both systems. In both cases, we assume we only have uncertainty on position, initialized as a unimodal distribution of particles. This distribution is updated online using a particle filter given noisy observations of distance from four laser range finder sensors aligned with cardinal directions in the vehicle frame. As the true observation noise is often not known exactly, we choose a different noise covariance for the particle filter based on performance of the downstream controller.
These particles are directly fed in to the SCP PMPC controller. For the CE controller, we plan assuming the state is the expected position of the particles.
In both settings, we evaluate performance over scenarios where the true position is sampled from initial belief over position.

Results for this setting are plotted in Figure \ref{fig:sensing}. Again, we see uniformly better performance for the PMPC controller compared to the CE controller. We note that while the costs of the PMPC and CE controller for the free-flyer appear close, the difference in cost is highly significant, since it is due to collisions occurring in the CE, but not the PMPC controller.

\begin{figure}
\centering
\subfigure[Free-flyer]{
\includegraphics[width=0.22\textwidth]{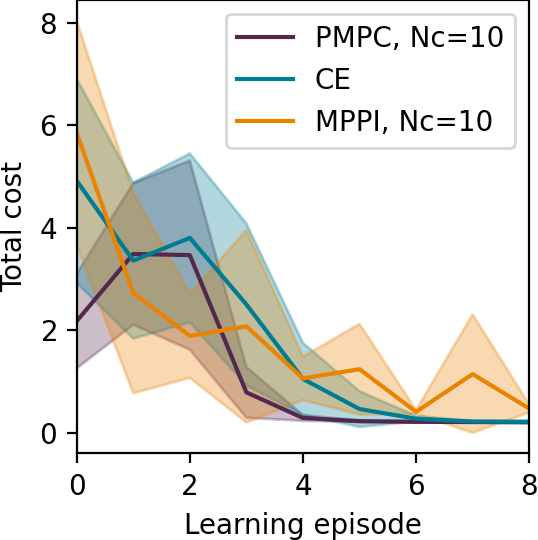}
}
\subfigure[Free-flyer]{
\includegraphics[width=0.22\textwidth]{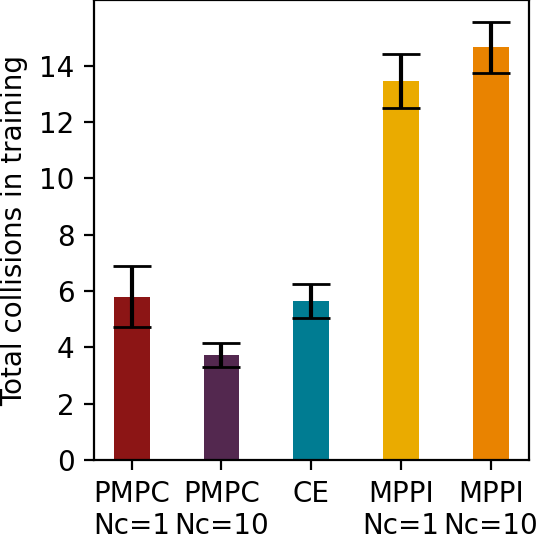}
}
\subfigure[Planar quadrotor]{
\includegraphics[width=0.22\textwidth]{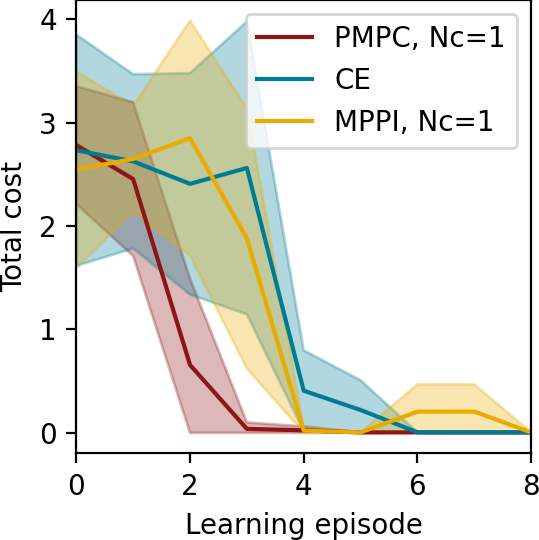}
}
\subfigure[Planar quadrotor]{
\includegraphics[width=0.22\textwidth]{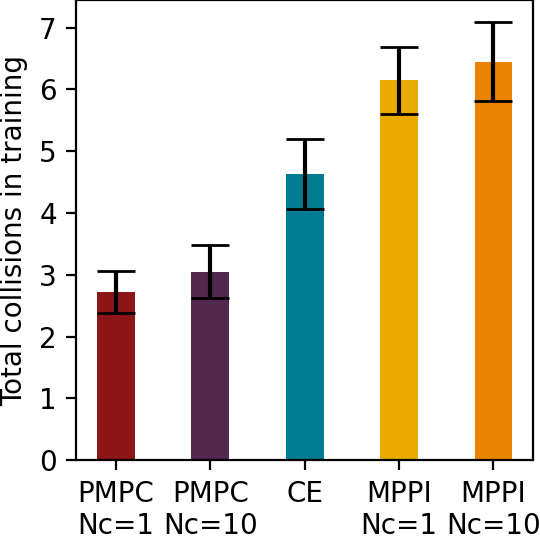}
}
\caption{Learning curves (cost per iteration) and total number of collisions during training for the free-flyer and the planar quadrotor in a model-based reinforcement learning setting.}
\label{fig:learning}
\end{figure}

\subsection{Learning for Control}
A third, and increasingly common source of uncertainty in robotics arises from learned components. Any learned component has \textit{epistemic uncertainty}, reflecting uncertainty in the underlying function as opposed to \textit{aleatoric}, or irreducible, uncertainty which is an intrinsic feature of the environment. Characterization of this epistemic uncertainty has led to substantial improvements in learning-based control and reinforcement learning due to better robustness with respect to unknown dynamics and better exploration \cite{chua_deep_2018}. A common tool to quantify this uncertainty is using deep ensembles \cite{lakshminarayanan2017simple}, wherein several neural networks are trained on the same data but starting from different initializations and regularized to different points. Following the problem setting of \cite{chua_deep_2018}, we consider the problem of control with a learned dynamics model. Specifically, we consider a model-based RL framework, as the initial low-data regime highlights the need to factor in epistemic uncertainty.

We train $D$ neural network dynamics models between every episode on the state, action, next-state tuples collected so far. We minimize a squared $\ell_2$ loss, with each network regularized to its random initialization, following methods proposed by \cite{osband2018randomized, pearce2020uncertainty}. 
During the episode, we use each network as the (stationary) dynamics for a single particle in the PMPC controller. We compare against a CE controller that uses a single network in the the ensembles (ignoring epistemic uncertainty) as well as an MPPI controller \cite{williams2017information} which factors in this uncertainty but is a sampling-based, gradient free method (as used in \cite{chua_deep_2018}).

Our results are visualized in Figure \ref{fig:learning}. Figures (a) and (c) show learning curves for the reinforcement learning process: they show the cost per episode during learning. There are several things to note in these figures: (i) both the CE and PMPC controller outperform the MPPI controller and (ii) the performance PMPC improves upon the CE method. (i) is likely the result of incorporating gradient-based information as opposed to using stochastic, sampling-based optimization schemes. (ii) shows the value of approximately incorporating uncertainty.
Although the collisions in training are relatively close, the training curves for both systems show convergence 2-4 episodes before the CE approach. Additionally, the MPPI controller shows performance variation in later episodes due to its stochastic nature, while both the CE and PMPC controllers achieve consistent performance.

\subsection{Computational Complexity}

\begin{wrapfigure}{R}{0.5\linewidth}
%\begin{figure}[H]
\begin{center}
\includegraphics[width=\linewidth]{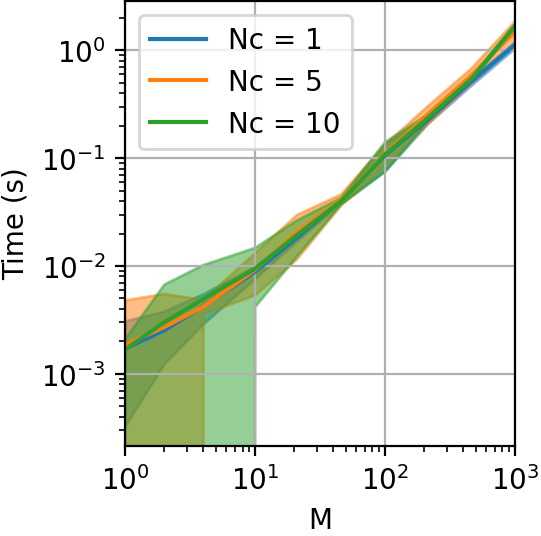}
\end{center}
\vspace{-20pt}
\caption{Time per SCP iteration of the nonlinear particle MPC scheme for varying number of particles, $M$.}
\label{fig:timing}
%\end{figure}
\end{wrapfigure}

Figure \ref{fig:timing} shows time per SCP iteration for a varying number of particles, for $\consensusHorz=1,5,10$. The plots were generated on a Intel(R) Core(TM) i7-8559U CPU @ 2.70GHz. The slope of this log-log plot is approximately 1, showing approximately linear computational complexity in the number of particles. This is due to the utilization of sparse quadratic programming solvers. Moreover, interestingly, the larger consensus period does not have a substantial impact on computational complexity until very large numbers of particles are used ($\approx1000$), where there is an approximately 1.5$\times$ performance difference for $\consensusHorz=5,10$ versus $\consensusHorz=1$. The results show for an intermediate number of particles (e.g. approximately 20), operational frequencies of around 50 are possible even with off-the-shelf solvers and standard consumer CPUs. With GPU acceleration, we anticipate operational frequencies of 100Hz with up to 50 particles is achievable in the near term. 

We highlight that the Monte Carlo methods we develop in this paper are currently becoming computationally feasible in real time due to advances in computational hardware for parallel processing and sparse solvers. Thus, we believe that effective design of Monte Carlo-based methods as opposed to imprecise approximations discussed in the introduction will be a fruitful avenue of research for uncertainty characterization in control in coming years.

%%%%%%%%%%%%%%%%%%%%%%%%%%%%%%%%%%%%%%%%%%%%%%%%%%%%%%%%%%%%%%%%%%%%%%%%%%%%%%%%
\section{Conclusions}

In this work we have presented a framework for particle-based nonlinear model predictive control enabling effective control under uncertainty, which offers tunable conservatism by using partial action consensus. We have demonstrated that on two different robot systems, accounting for three different uncertainties common in robotics, PMPC outperforms a certainty equivalent baseline.
In the model-based RL (MBRL) problem setting, PMPC substantially outperforms uncertainty-aware MPPI \cite{williams2017information}, which has been shown to substantially improve MBRL relative to CE methods \cite{chua_deep_2018}. 
Finally, we have investigated the computational complexity of the proposed approach and found that the method is capable of real time operation, and will become increasingly capable for dynamic real time applications as computational hardware and sparse solvers improve. Thus, further algorithmic improvements in particle-based control methods represent a promising direction of future research for control under uncertainty.

\nocite{*}

\bibliography{Biblio}

%%%%%%%%%%%%%%%%%%%%%%%%%%%%%%%%%%%%%%%%%%%%%%%%%%%%%%%%%%%%%%%%%%%%%%%%%%%%%%%%
\clearpage
\section*{Appendix}

\subsection{Proof of Theorem 1}

We restate theorem 1 for completeness:

\setcounter{theorem}{0}
\begin{theorem}[Consensus horizon \& replanning]
\label{thm:consensus}
For any integers $N^\star \in [2..\infty)$, $K \in [1..N^\star-1]$, there exists an optimal control problem \eqref{eqn:fhoptcontrol} defined by $\xdist,f,c,\mathbb{X},\mathbb{U},\mathbb{W}$ for which running PMPC with a consensus horizon of $\consensusHorz = K < N^\star$ leads to constraint violation, but a consensus horizon of $\consensusHorz = N^\star$ does not.

\end{theorem}

\begin{proof}
Define an uncertain bimodal system $f$ 
\begin{align*}
x^{(0)} \sim \xdist &= \{ 0 \}\\
\mathbb{X} &= \{0, 1\}\\
\mathbb{U} &= \{0, 1 \}\\
\mathbb{W} &= \{ 0, 1 \}
\end{align*}
where transitions $\xjp = f(\xj, \uj, \wj)$ and the state dependent cost are 
\begin{align*}
f(\xj, \uj, \wj) &= \begin{cases} 
\begin{cases}
0 \text{ if } \uj = 0 \\
1 \text{ if } \uj = 1
\end{cases} & \text{ if } \xj = 0 \\
\begin{cases}
1 \text{ if } \uj = \wj \\
2 \text{ if } \uj = \neg \wj
\end{cases} & \text{ if } \xj = 1 \\
\begin{cases}
2 \text{ if } \uj = 0 \\
2 \text{ if } \uj = 1
\end{cases}& \text{ if } \xj = 2 
\end{cases}
\end{align*}
\begin{align*}
c(\xj) &= \begin{cases}
1 & \text{ if } \xj = 0\\
0 & \text{ if } \xj = 1\\
\infty \text{ (\emph{constraint violation})} & \text{ if } \xj = 2\\
\end{cases}
\end{align*}
Note that $\xj = 2 \notin \mathbb{X}$ implies a constraint violation.

For the bimodal parameter uncertainty distribution between $w_1$ or $w_2$
\begin{align*}
w_1^{(j)} &= \begin{cases}
0 & \text{ if } j < N^\star \\
j \bmod 2 & \text{ if } j \geq N^\star
\end{cases} \\
w_2^{(j)} &= \begin{cases}
0 & \text{ if } j < N^\star \\
(j + 1) \bmod 2 & \text{ if } j \geq N^\star
\end{cases}
\end{align*}
For receding planning at time $j = 0$, the control consensus of $N^\star$ suggests a control policy with $u^{(j)} = 0$ which \emph{does not} lead to constraint violation.

Control consensus of $K < N^\star$ suggests a policy with $u^{(0)} = 1$ which leads to constraint violation, because too fast information gain is assumed, whereas in reality the system cannot be stabilized at $\xj = 1$.

This shows that for any integers $N^\star \in [2..\infty)$, $K \in [1..N^\star-1]$, there exists an optimal control problem \eqref{eqn:fhoptcontrol} with uncertainty such that running PMPC with consensus horizon $K$ leads to constraint violation, but a consensus horizon of $N^\star$ correctly accounts for uncertainty, avoiding constraint violation.

\end{proof}

\subsection{Experimental Details}

\begin{figure}
\centering
\subfigure[Free-flyer, state uncertainty]{
\includegraphics[width=0.22\textwidth]{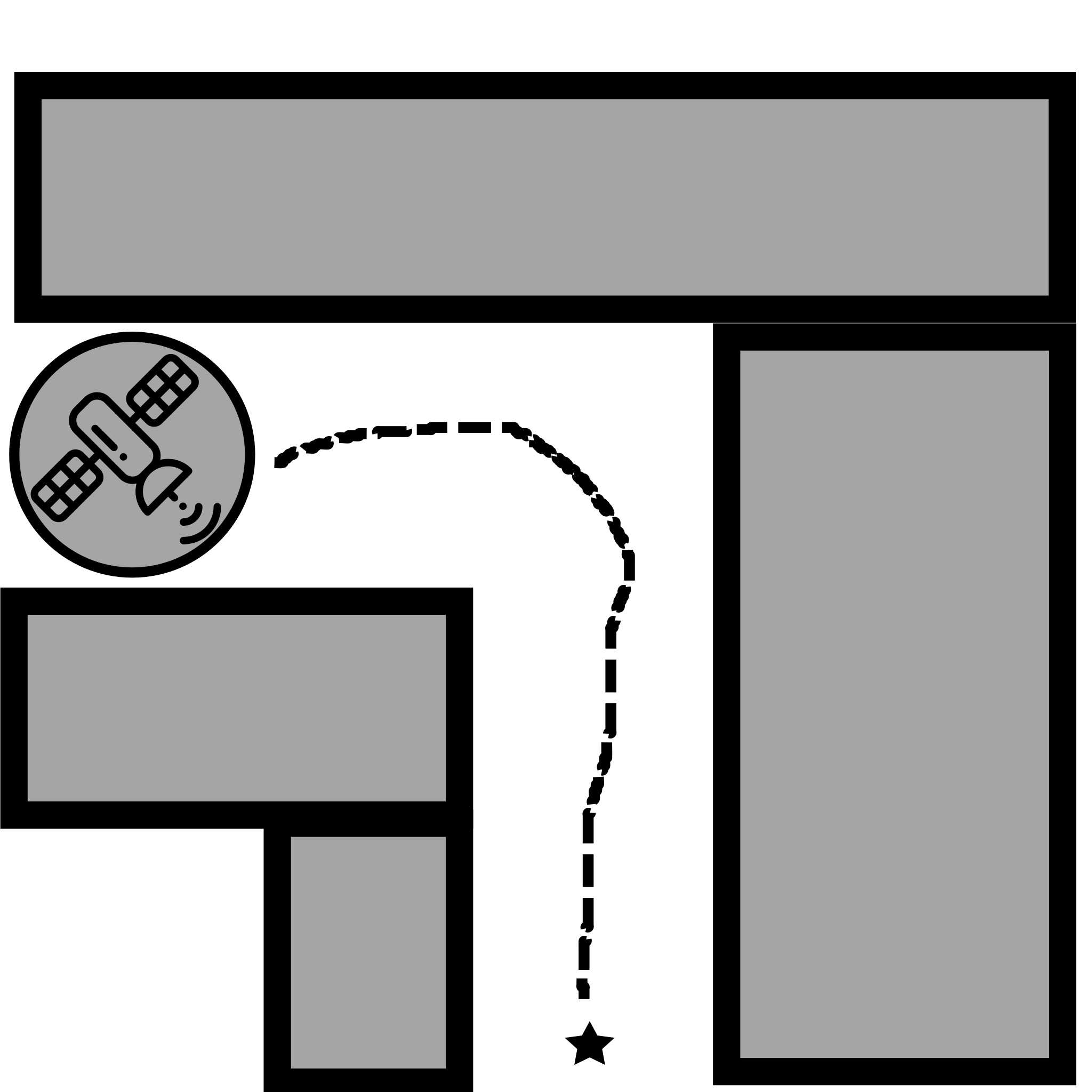}
}
\subfigure[Free-flyer, model uncertainty]{
\includegraphics[width=0.22\textwidth]{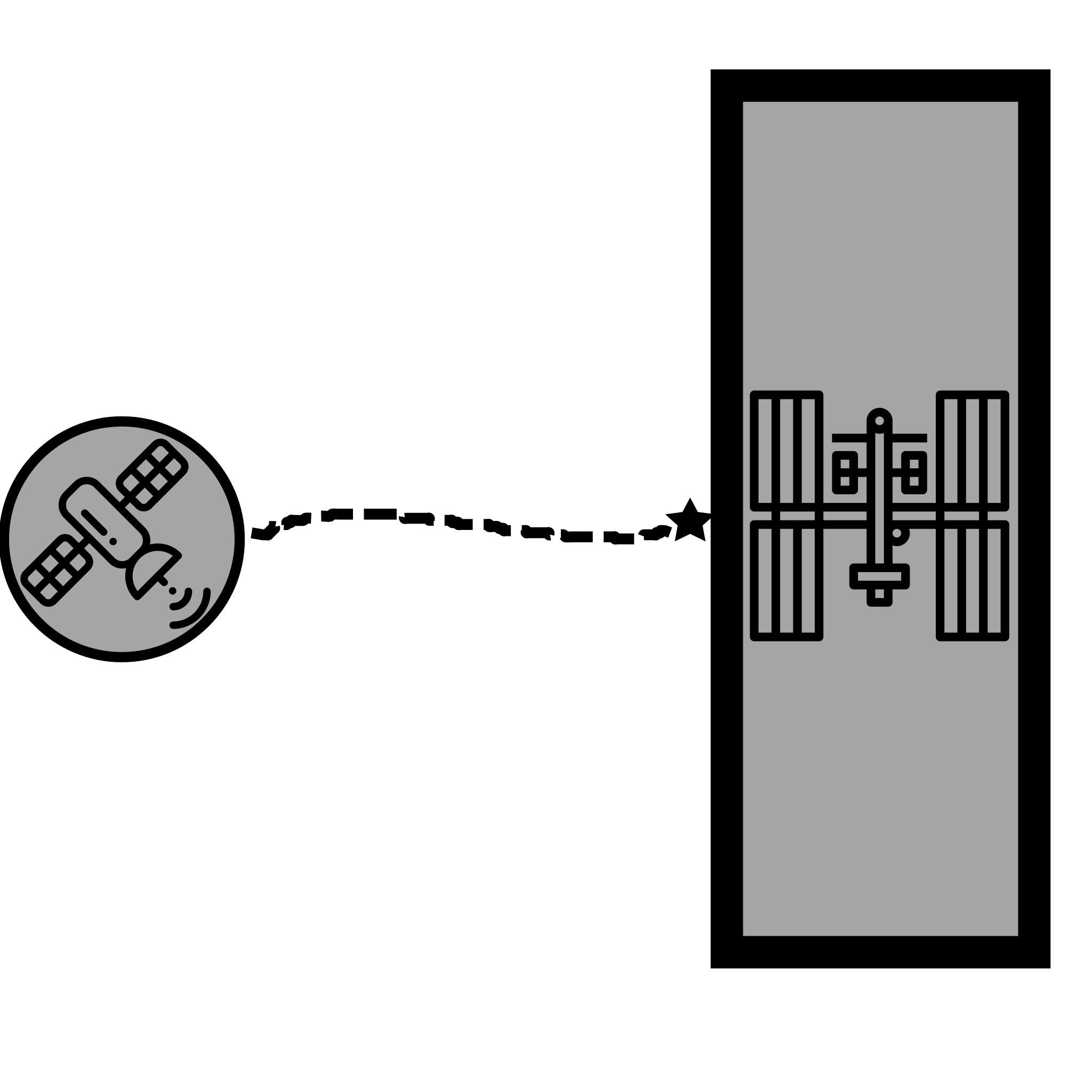}
}
\subfigure[Planar quadrotor, state uncertainty]{
\includegraphics[width=0.22\textwidth]{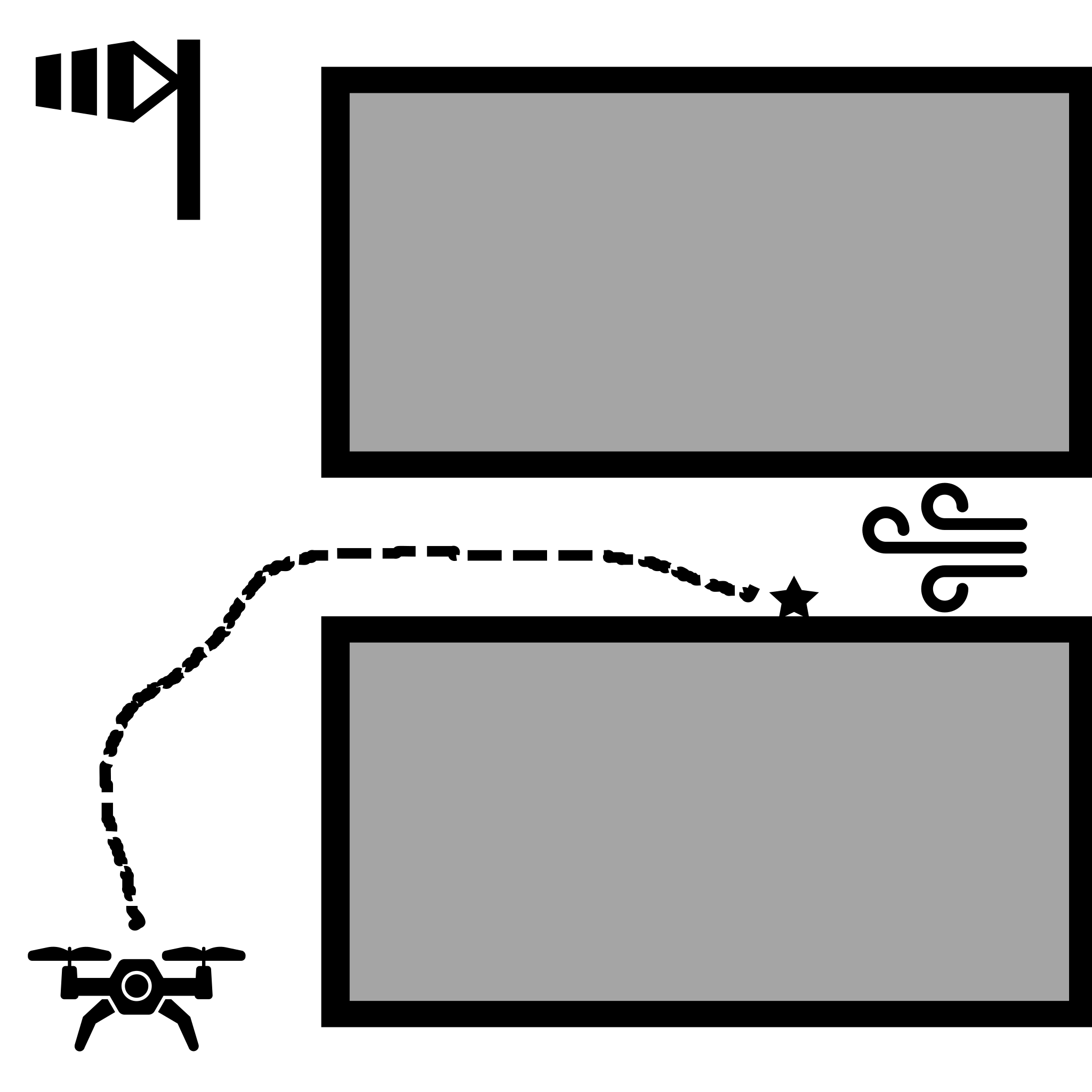}
}
\subfigure[Planar quadrotor, model uncertainty]{
\includegraphics[width=0.22\textwidth]{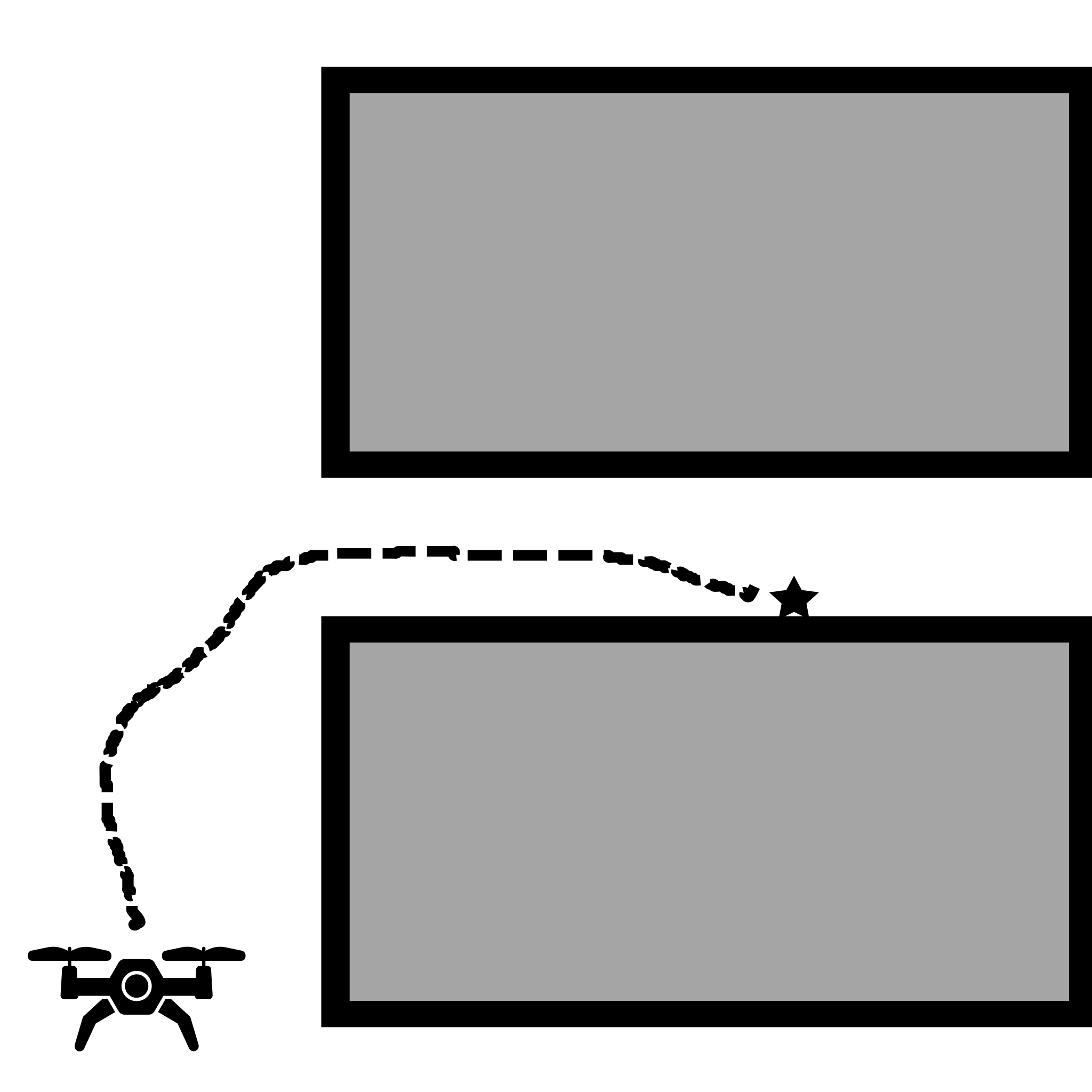}
}
\caption{Experimental setup sketch}
\label{fig:envs}
\end{figure}

\subsubsection{Obstacles Implementation}

In all experiments we enforce no obstacle violation via exact penalty functions in the objective. Such penalty specification has the property that it is enforced exactly for a sufficiently large penalty weight. All obstacles considered in the experiments are rectangles and, for the penalty function, we pick the distance to the nearest edge multiplied by a positive weight. We choose penalty approach instead of enforcing hard constraints to ensure plan feasibility: (i) during SCP convergence; (ii) in particle consensus planning, where a large number of different dynamics can include plans that cannot avoid obstacles. We choose the exact penalty approach because it results in the hard constraint solution at optimality and when final plan feasibility is possible.

\subsubsection{MPPI Implementation}

We implement MPPI following \cite{abraham2020model} where we take the weighting over control samples and sampled models. Since we use a consensus horizon shorter than the planning horizon, unlike \cite{abraham2020model}, we weigh actions \emph{separately for each model} past the consensus horizon. We attempted to automatically tune the MPPI parameters of (i) the number of iterations per time step, (ii) the number of random action sequences per iteration, (iii) the standard deviation of the normal sampling, (iv) the temperature of the algorithm. In both environments we choose these parameters by grid search to establish those which produced the lowest cost trajectories.

\subsubsection{Dynamics Uncertainty}

\paragraph{Free-flyer.}
We consider a setting in which the free-flyer must move to a location near a wall, but at each timestep, any of its four thrusters can fail with a probability of 5\%. 

Once an actuator fails, we assume it never recovers. In this setting, there are $2^4=16$ possible dynamics models corresponding to every combination of working and failed thrusters. Furthermore, dynamics are nonstationary, as we can probabilistically transition from working states to failure states over the course of time.

We assume the actuator failure is not directly observable and use a recursive Bayes filter to detect actuator failure. This filter maintains a belief over each of the $16$ possible states of the system. For PMPC, each particle must represent one (time-varying) dynamics model. To obtain each particle $i$, we simulate the evolution of dynamics over time according to the probability of failure at each time step from the current state to obtain a time-varying dynamics function for each particle. We repeat this multiple times for each hypothesis in the particle filter to obtain $M$ particles for PMPC. Each particle is weighted by the weight assigned to the root state by the particle filter. We compare to a certainty equivalent nonlinear MPC controller, in which we simulate actuator failure forward in time from the current state, and then select the maximum a posteriori (MAP) estimate for each time step. For the sake of comparison, we consider an oracle model which plans with access to the true dynamics evolution. All scenarios use a planning horizon $N = 20$, and PMPC uses a consensus horizon of $N_\text{consensus} = 7$ which was found to perform best. 

The starting position is at $(x = -20, y = 0)$ and zero translational and angular velocity. The wall occupies $x > 0.0$ and its exact penalty weight was $10^2$.

\paragraph{Planar quadrotor.}
For the planar quadrotor setting, we consider wind disturbances that evolve according to an Ornstein-Uhlenbeck (OU) process independently in the $x$ and $y$ directions:
\[
v_\text{wind}^{(j+1)} = \alpha v_\text{wind}^{(j)} + \epsilon \quad\quad
\alpha = 0.9 \quad\quad
\epsilon \sim \mathcal{N}(0, 2.0).
\]
The goal is to fly the quadrotor into a narrow passageway, defined by $(x > 0.0, y \in [10, 13])$, and regulate to the lower wall at $(x=15, y=10)$ in the presence of these noisy, changing wind disturbances. 

Here we assume the current wind is fully observed, so we do not require a particle filter. However, there is uncertainty over the future evolution of the wind, which we incorporate into PMPC. To sample the particles, we sample $M$ trajectories of the OU process. Each of these define a particle of time-varying dynamics $\{ \fj_i \}_{j=1}^N$. For a CE baseline, we consider planning assuming the current wind stays constant. Again, we compare to an oracle which knows the exact evolution of wind dynamics. We use a planning horizon of $N=20$, $M=10$ particles, and vary $\consensusHorz \in \{1, 5, 10\}$.

In all quadrotor settings the passageway is implemented by two obstacles, one above and one below with a penalty weight of $10^3$. The starting position is at zero translation and angular velocity at $(x=-3, y=12)$, to the left and slightly below the entrance to the passageway.

\subsubsection{State Uncertainty}

\paragraph{Free-flyer.}
\label{sec:ff_sense}
For the free-flyer, the task is to navigate around a corner in a narrow corridor. We use a planning horizon of $N=20$, $M=10$ particles, and a consensus horizon of $\consensusHorz=5$ in PMPC. We see that PMPC in this setting does better than CE, leading to a lower chance of collision and overall lower cost.

We consider uncertainty only in the 2D position. To do so, at the beginning of each episode, we sample a fixed point cloud of position deviations with a standard deviation of 3 m; we maintain the same deviation points throughout the episode. The true position has zero deviation and is included as one of the particles in the point cloud. 

Consequently, we use recursive Bayes filtering to estimate the true position. The free-flyer is equipped with 4 distance sensors along its two local axes, which obtain a noisy measurement of position with standard deviation of 3 m. In the filter, we assume a larger standard deviation of the observation noise of 10 m both because using the true observation noise led to near instantaneous adaptation and because the observation noise is often not known in practice. Using smaller than true observation noise can lead to spurious convergence to the wrong particle estimate and is not advisable.

In all free-flyer experiments the starting position is at $(x = 0, y = 0)$ and zero translational and angular velocity. The free space in the corridor spans $y = \in [-5, 5]$ meters vertically for $x < 20$ and $x \in [10, 20]$ for $y < 5$ to form an L-shaped corridor. The goal state is at $(x=15, y=-20$; in the middle of the vertical corridor part. There are two obstacles outer to the L-turn with a penalty weight of $10^2$ and one inner to the turn with a weight of $10^3$.

\paragraph{Planar quadrotor.}
For the planar quadrotor, the goal is to fly the quadrotor into a narrow passageway, defined by $(x > 0.0, y \in [10, 13])$, and regulate to the lower wall at $(x=15, y=10)$ in the presence of only position uncertainty. To do so, at the beginning of each episode, we sample a fixed point cloud of position deviations with a standard deviation of 1 m; we maintain the same deviation points throughout the episode. The true position has zero deviation and is included as one of the particles in the point cloud. 

Consequently, we use a recursive Bayes filter to determine the true deviation. The free-flyer is equipped with 4 distance sensors along its two local axes, which obtain a noisy measurement of position with a standard deviation of 1 m. In the filter, we use the true observation noise. The adaptation with true noise is much slower than for the free-flyer in a corridor, likely because the quadrotor initially only has a wall to its right, unlike the free-flyer which typically registers walls on all of its sides and so has more measurements available.

In all quadrotor settings the passageway is implemented by two obstacles, one above and one below with a penalty weight of $10^3$. The starting position is at zero translation and angular velocity at $(x=-3, y=5)$, to the left and below the entrance to the passageway.

\subsubsection{Learning for Control}

For both environments we learn by fitting an ensemble network model of the dynamics after each episode---we train D neural networks the weights of each are regularized to a random initialization using weighted squared $\ell_2$ distance to capture epistemic uncertainty. For the CE case we train a single un-regularized neural network with the same architecture. We use a fixed learning rate of $10^{-3}$ and train until convergence on the noise-free dynamical transitions data obtained in all previous episodes.

\paragraph{Free-flyer.}

We test the free-flyer in the same corridor navigation task as in \ref{sec:ff_sense}. We use a 3-layer feed-forward neural network model with 3 hidden layers of width 64. For the ensemble network we regularize the weights by $0.001$ divided by the output dimension; we do this to capture epistemic uncertainty. The angle component of the input is processed into $(\sin(\theta), \cos(\theta))$ before being fed to the network to prevent wraparound issues. We use $M = 10$ particles, a planning horizon of $N = 20$, a consensus horizon of $N_\text{consensus} = 10$. We introduce normal control noise to all control strategies for the first $6$ episodes to induce exploration, with a decaying sequence of standard deviations $(0.3, 0.3, 0.1, 0.1, 0.01, 0.01)$. 
The starting and goal position and the obstacles are identical to the one in \ref{sec:ff_sense}.
For the MPPI implementation we use 50 iterations per timestep, 60 action sequences, a standard deviation for sampling of $0.3$ and a temperature of $10^{-3}$.

\paragraph{Planar quadrotor.}
The planar quadrotor is tested in the same narrow passage way task, but here without wind. The architecture of the network is identical to that for the free-flyer except of the width of 32. As learning happens faster for this environment, we found we only needed to add exploration control noise to the first 3 episodes with a standard deviation sequence of $(0.3, 0.1, 0.01)$. All controller parameters are identical to the free-flyer setup, except here we use $\consensusHorz=1$. 

In all quadrotor settings the passageway is implemented by two obstacles, one above and one below with a penalty weight of $10^3$. The starting position is at zero translation and angular velocity at $(x=0, y=0)$, slightly to the left and significantly below the entrance to the passageway.
For the MPPI implementation we use 30 iterations per timestep, 100 action sequences, a standard deviation for sampling of $0.2$ and a temperature of $10^{-5}$.

\end{document}